\documentclass[aps,preprint,prd]{revtex4-1}

\usepackage{amsmath,amssymb,amsfonts,graphicx,hyperref,amsthm,color}

\usepackage{epsfig}
\usepackage{float}
\usepackage{multirow}
\usepackage{subfig}

\newcommand{\beq}{\begin{equation}}
\newcommand{\eeq}{\end{equation}}
\newcommand{\bea}{\begin{eqnarray}}
\newcommand{\eea}{\end{eqnarray}}

\begin{document}

\title{Rainbow's Gravity Corrections to the Black Hole Global Casimir Effect}

\author{G. Alencar}
\email{geova@fisica.ufc.br}\address{Departamento de F\'{\i}sica, Universidade Federal do Cear\'{a},
Caixa Postal 6030, Campus do Pici, CEP 60455-760, Fortaleza, Cear\'{a}, Brazil.}

\author{R.N. Costa Filho}
\email{rai@fisica.ufc.br}\address{Departamento de F\'{\i}sica, Universidade Federal do Cear\'{a},
Caixa Postal 6030, Campus do Pici, CEP 60455-760, Fortaleza, Cear\'{a}, Brazil.}

\author{M.S. Cunha }
\email{marcony.cunha@uece.br}\address{Grupo de F\'isica Te\'orica (GFT), Centro de Ci\^encias e Tecnologia, Universidade Estadual do Cear\'a, CEP 60714-903, Fortaleza, Cear\'a, Brazil.}

\author{C.R. Muniz}
\email{celio.muniz@uece.br}\address{Grupo de F\'isica Te\'orica (GFT), Universidade Estadual do Cear\'a, Faculdade de Educa\c c\~ao, Ci\^encias e Letras de Iguatu, CEP 63900-000, Iguatu, Cear\'a, Brazil.}

\begin{abstract}
In this manuscript we compute corrections to the global Casimir effect at zero and finite temperature due to Rainbow's Gravity (parametrized by $\xi$). For this we use the solutions for the scalar field with mass $m$ in the deformed Schwarzschild background and the corresponding quantized energies of the system, which represent the stationary states of the field and yield the stable part of the quantum vacuum energy. The analysis is made here by considering the limit for which the source mass, $M$, approaches zero, in order to verify the effects on the global Casimir effect in mini black holes near to the Planck scale, $\omega_P$. We find a singular behavior for the regularized vacuum energy at zero temperature and for all the corresponding thermodynamic quantities when $m^2=\omega^2_P/\xi$, what can be seen as the limit of validity of the model. Furthermore, we show that the remnant Casimir tension over the event horizon in the limit $M\to 0$ is finite for any temperature and all the space of parameters. In fact we show that the remnant tension receives no corrections from Rainbow's Gravity. This points to the fact that such a behavior may be an universal property of this kind of system.

  \vspace{0.75cm}
\end{abstract}

\maketitle

\section{Introduction}

One of the major goals of theoretical physics nowadays is to discover a full quantum theory for the gravitational interaction, which would include backreaction effects of the quantum fields on the spacetime geometry. Unfortunately, the best theory available so far is still the old quantum field theory in curved spacetimes, which is semiclassical in the sense of admitting quantum fields living in a classical smooth spacetime background, without backreaction. Attempts to go beyond the semiclassical theories and to approximate to the quantum gravity consider, in several approaches, the Lorentz symmetry breaking in high energies (UV regimen). In such theoretical descriptions the spacetime effects on the quantum fields are quite nontrivial, as, for instance, in the Ho\v{r}ava-Lifshitz theory (H-L) \cite{Horava:2009uw,Horava:2009if,Alencar:2015aea,Muniz:2014dga,Bernardini:2017jlz}.

Rainbow's gravity (RG) is another effective quantum gravity theory that implies the Lorentz symmetry breaking in the UV regimen. This theory can be obtained as a generalization to curved spacetimes of the Doubly Special Relativity (DSR) \cite{Girelli:2004ms,Imilkowska:2005vs,AmelinoCamelia:2005ne,Hackett:2005mb,AmelinoCamelia:2004pv}. In this latter, the energy-momentum dispersion relations become nonlinear in very high energies, presupposing also a maximal energy $\omega_{max}$ or a minimal length $\ell_{min}$. Such quantity is explicitly included in those relations, leading to a theory fundamentally non-local whereas such magnitudes are the Planckian ones (e.g. $\omega_{max}=\omega_P$, $\ell_{min}=\ell_P$), since it can be derived from the non-commutative geometry \cite{Garattini:2013yha}. The DSR theory can also be seen as having a ``deformed'' Lorentz symmetry \cite{Amati:1988tn}, such that the standard energy-momentum relations (MDR) in flat spacetime are modified by Planck scale corrections involving functions $f(\omega/\omega_P)$ and $g(\omega/\omega_P)$ such that
\begin{equation}\label{MDR}
\omega^2\,f^2(\omega/\omega_P)-p^2\,g^2(\omega/\omega_P)=m^2 c^4,
\end{equation}
where $m$ is the rest frame mass of the particle. Global Lorentz invariance is in fact an accidental symmetry related to a particular solution of General Relativity. It is evident that, to be consistent with the standard theory, the functions $f(\omega/\omega_P)$ and $g(\omega/\omega_P)$  which appear in Eq. (\ref{MDR}) must tend to unity for $\omega\ll\omega_P$. It is interesting to note that in some cases there is a correspondence between the H-L and RG descriptions \cite{Garattini:2014rwa}.

In this context, curved spacetimes are now endowed with a nontrivial quadratic invariant, namely, an energy-dependent metric tensor. It means that if a certain observer measures a particle (or wave) with energy $\omega$, then he or she concludes that this probe perceives a metric that generally depends on $\omega/\omega_P$, which can be seen as a kind of backreaction of the fields on the spacetime. As well as in the H-L theory, the so deformed background spacetime induces nontrivial effects on the quantum fields when compared with their merely semiclassical counterparts \cite{Ali:2014aba,Garattini:2017bgx,Bezerra:2017hrb,Gim:2017rmn,Heydarzade:2017rpb,Nilsson:2016rsv}, effects which can be local (geometric) and/or global (topological), as in the gravitational Casimir effect \cite{Zhuk:1996xc,Bezerra:2017zqq}.

The Casimir effect is a phenomenon associated to the quantum vacuum oscillations of quantum fields, and originally was studied from the modifications, in Minkowski spacetime, of the zero point oscillations of the electromagnetic field due to the presence of material boundaries, at zero temperature \cite{Casimir:1948dh}. The present status of the phenomenon is that the effect can also occur considering other quantum fields
(for a contemporary review, see \cite{Bordag:2009zzd}), and also due to nontrivial topologies associated with different spacetimes \cite{Bezerra:2011nc,Bezerra:2014pza,Mota:2015ppk,Bezerra:2016qof} and even with quantum spaces linked to condensed matter systems \cite{Muniz:2014pha}.

The investigation of the Casimir effect caused by the non-trivial topology of the spacetime, and not by the presence of physical boundaries, in obtaining the regularized physical quantities associated with the quantum vacuum can unconsider the computation of local magnitudes (as the vacuum expected value of the momentum-energy density) just due to the fact of being a global or topological effect. Thus, the regularization of these quantities can be made directly via Zeta functions \cite{Elizalde:1994gf}. In fact, this aspect was recently explored in an article \cite{Muniz:2015jba} that studied the effects of the stable part of the quantum vacuum of the massive scalar field around a static black hole, including the thermal effects in mini black holes. Their authors considered the black hole as having a nontrivial topology due to existence of the singularity, without considering the presence of material boundaries.

The present paper seeks to make an extension of the aforementioned work in order to include aspects of quantum gravity through the deformation of the Schwarzschild spacetime via introduction of the MDR functions defined as
\beq  \label{RainbowFunction}
f(\omega/\omega_P)=1, \ \ \ \ \ \ g(\omega/\omega_P)=\sqrt{1-\xi(\omega/\omega_P)^s},
\eeq
where $\xi>0$, and $s$ is a positive integer of order one. These functions are inspired in loop quantum gravity theory and non-commutative space models of gravity \cite{Smolin:2005cz}. Thus, the present work goes in the direction of investigating, besides the topological Casimir effect around the black hole, the non-local influences due to the Rainbow's Gravity (parameterized here by $\omega_P=1/\ell_P$ and $\xi$). The work will focus on the mini black holes, where the quantum gravity effects are arguably more pronounceable and where the quantum vacuum can be decisive, in their final moments \cite{Zeldovich:1981wc}.

The paper is organized as follows: In section II we study the global Casimir effect at zero temperature, in section III we proceed with the analysis at finite temperature and finally, in section IV, we draw the conclusions.

\section{Global Casimir Effect in the Rainbow's Gravity at Zero Temperature}

Prior to analyse the global Casimir effect, we must review here the solutions of the Rainbow's Gravity (RG) covariant Klein-Gordon equation of massive scalars minimally coupled to the Schwarzschild gravitational field, given by \cite{Bezerra:2017hrb}
\begin{equation}
\left[\frac{1}{\sqrt{-\mathrm{g}}}\partial_{\mu}\left(\mathrm{g}^{\mu\nu}\sqrt{-\mathrm{g}}
\partial_{\nu}\right)+m^{2}\right]\Psi=0\ ,
\label{eq:Klein-Gordon_cova}
\end{equation}
(where natural units $c \equiv \hbar \equiv 1$ are used).

The gravitational background generated by a static uncharged compact object is given
by the Schwarzschild metric now depending on the MDR functions $f(\omega/\omega_P)$ and $g(\omega/\omega_P)$.
In spherical coordinates the square line-element invariant reads
\begin{equation}
ds^{2}=f^{-2}(\omega/\omega_P)h(r)dt^{2}-g^{-2}(\omega/\omega_P)[h(r)^{-1}dr^{2}+r^{2}d\Omega^2],
\label{eq:metrica_Kerr-Newman}
\end{equation}
where $h(r)=\left(1-\frac{r_s}{r}\right)$, $r_s=2MG$ is the Schwarzschild radius,
 $d\Omega^2=d\theta^{2}+\sin^{2}\theta\ d\phi^{2}$, $G=G(0)$
is the Newton's universal gravitational constant and $M$ is the mass of the source.
By symmetry arguments we assume that solutions of Eq.~(\ref{eq:Klein-Gordon_cova}) can be factored as follows
\begin{equation}
\Psi(\mathbf{r},t)=R(r)Y_{l}^{m_l}(\theta,\phi)\mbox{e}^{-i\omega t},
\label{eq:separacao_variaveis}
\end{equation}
where $Y_{l}^{m_l}(\theta,\phi)$ are the spherical harmonic
 functions. Inserting Eq. (\ref{eq:separacao_variaveis})
and the metric given by Eq.~(\ref{eq:metrica_Kerr-Newman}) into (\ref{eq:Klein-Gordon_cova}),
we obtain the following radial equation
\begin{equation}
\frac{d}{dr}\left[r(r-2GM)\frac{dR}{dr}\right]
+\left(\frac{r^{3}\tilde{\omega}^{2}}{r-2GM}-\tilde{m}^{2}r^{2}-\lambda_{lm_l}\right)R=0,
\label{eq:mov_radial_1}
\end{equation}
where $\lambda_{lm_l}=l(l+1)$ and
\begin{eqnarray}\label{tildes}
 \tilde{\omega}&=&\frac{f(\omega/\omega_P)}{g(\omega/\omega_P)}\omega, \\ \nonumber
 \tilde{m}&=&\frac{m}{g(\omega/\omega_P)}.
 \end{eqnarray}

The general solution to Eq.~(\ref{eq:mov_radial_1}) is given in terms of the confluent Heun's functions
\begin{eqnarray}
R(x) & = &C_{1} \mbox{e}^{\frac{1}{2}\alpha x}x^{\frac{1}{2}\beta} \mbox{HeunC}(\alpha,\beta,\gamma,\delta,\eta;x)\nonumber\\
&+&C_{2}\mbox{e}^{\frac{1}{2}\alpha x}\ x^{-\frac{1}{2}\beta}\mbox{HeunC}(\alpha,-\beta,\gamma,\delta,\eta;x)\ ,
\label{eq:solucao_geral_radial_Kerr-Newman_gauge}
\end{eqnarray}
where $x=(r-2GM)/2GM$, over the entire range $0<x\leq \infty$, with $C_{1}$ and $C_{2}$ being constants, and the parameters $\alpha$, $\beta$, $\gamma$, $\delta$, and $\eta$ are explicitly written in terms of the rainbow's function are given by:
\begin{subequations}

\begin{equation}
\alpha=-4M\frac{\sqrt{m^{2}-\omega^{2}f^2(\omega/\omega_P)}}{g(\omega/\omega_P)} ;
\label{eq:alpha_radial_HeunC_Kerr-Newman}
\end{equation}
\begin{equation}
\beta=\frac{i4M\omega}{g(\omega/\omega_P)}\ ;
\label{eq:beta_radial_HeunC_Kerr-Newman}
\end{equation}
\begin{equation}
\gamma=0\ ;
\label{eq:gamma_radial_HeunC_Kerr-Newman}
\end{equation}
\begin{equation}
\delta=\frac{4M^{2}\left[m^{2}-2\omega^{2}f^2(\omega/\omega_P)\right]}{g^2(\omega/\omega_P)};
\label{eq:delta_radial_HeunC_Kerr-Newman}
\end{equation}
\begin{equation}
\eta=-l(l+1)-\frac{4M^{2}\left[m^{2}-2\omega^{2}f^2(\omega/\omega_P)\right]}{g^2(\omega/\omega_P)}.
\label{eq:eta_radial_HeunC_Kerr-Newman}
\end{equation}

\end{subequations}

This is the sum of two linearly independent solutions of the confluent Heun differential
equation provided $\beta$ is not an integer.

In order to have a polynomial confluent Heun function
(\ref{eq:solucao_geral_radial_Kerr-Newman_gauge}),
we must impose the so called $\delta_N$ and $\Delta_{N+1}$
conditions
\begin{eqnarray}
\frac{\delta}{\alpha}+\frac{\beta+\gamma}{2}+1&=&-N \label{eq:cond_polin_1}\\
\Delta_{N+1}&=&0
\label{eq:cond_polin_2}
\end{eqnarray}
where $N$ is a non-negative integer.
From Eqs. (\ref{eq:cond_polin_1}) and (\ref{RainbowFunction}) (with $s=2$), we obtain the following expression for the energy levels
\begin{equation}
n+g^{-1}(\omega/\omega_P)\left[2iM\omega-\frac{M\left(m^{2}-2\omega^{2}\right)}{\sqrt{m^{2}-\omega^{2}}}\right]=0,
\label{eq:energy_levels}
\end{equation}
where $n=N+1$.

Now we must compute the corrections to the global Casimir effect due to the RG. Just as in Ref. \cite{Muniz:2015jba} we consider two cases: at zero and at finite temperature. Here we consider the first one. For both cases we will need of an explicit expression for the energy modes of the system. In order to obtain this we have to take the limit $M\omega \to 0$ in Eq. (\ref{eq:energy_levels}). This is a natural cutoff since black holes there do not absorb waves when $M\omega \lesssim1/\sqrt{54}$ \cite{Nugaev:1979fj}. Choosing $s = 1$ in Eq. (\ref{RainbowFunction}) leads us to two complex omega solutions and a real one but this diverges as $\xi\to 0$. For $s = 2$ there is only one relevant solution and we get therefore the corrected energy for the massive scalar field in the RG, which is given by
\begin{equation}\label{massiveenergies}
\omega_{n}^{2}=\frac{m^{2}}{2}+\frac{1}{2\xi}-\frac{\sqrt{m^{4}\xi^{2}-2m^{2}\xi-8m\xi\omega_{n}^{(0)}+1}}{2\xi}
\end{equation}
where $\omega_{n}^{(0)}=-m^{3}M^{2}/2n^{2}$. Since we are interested in the computation of a remnant Casimir tension we must consider an expansion in the physical parameter $M$. The modification due to the RG to the energy of the system is then given by
\begin{equation}\label{massiveenergies xi}
\omega_n =\left(\frac{1+\xi m^{2}-|1-\xi m^{2}|}{2\xi}\right)^{\frac{1}{2}}-\frac{m^{4}\sqrt{\xi}|1-m^2\xi|^{-1}}{n^2\sqrt{2}\sqrt{1+m^{2}\xi-|1-m^{2}\xi|}}M^2+\mathcal{O}(M^4).
\end{equation}

We should point that when we take $\xi\to 0$ the above expression recovers the one found in Ref \cite{Muniz:2015jba} for the case without RG. For $\xi\neq 0$ the right hand side of Eq. (\ref{massiveenergies xi}) above is singular when $m^2\xi=1$. If we recover the unities, we can find that in fact the singularity happens at  $m\approx\omega_P/\sqrt{\xi}$. Therefore this singularity just point to the sub-Planckian limit of validity of the system. We will see that this singularity will also be present in most of the regularized quantities studied here. Anyway we also study the behavior of the system in both regimes: the sub-Planckian with $m^2\xi<1$ and the trans-Planckian with $m^2\xi>1$. In these regimes Eq. (\ref{massiveenergies xi}) simplifies to
\begin{equation}\label{m<1}
\omega_n \approx \begin{cases} m-\frac{m^{3}M^2}{2n^2(1-m^2\xi)} &\mbox{for } m^2\xi<1, \\
\frac{1}{\sqrt{\xi}}-\frac{m^{4}\sqrt{\xi}}{2n^2(m^2\xi-1)}M^2 & \mbox{for } m^2\xi>1. \end{cases}
\end{equation}

Now we compute the corrections to the remnant Casimir tension due to the RG at zero temperature. First, by using Eq. (\ref{m<1}), we can compute the quantum non-regularized vacuum energy
\begin{equation}\label{vacuumenergydef}
E^{(0)}=\frac{1}{2}\sum_n n^{2}\omega_{n}
\end{equation}
which gives us
\begin{equation}
E^{(0)} \approx \begin{cases} \frac{1}{2}\sum_n n^{2}\left(m-\frac{m^{3}M^2}{2n^2(1-m^2\xi)}\right)   &\mbox {for } m^2\xi<1, \\
\frac{1}{2}\sum_n n^{2}\left(\frac{1}{\sqrt{\xi}}-\frac{m^{4}\sqrt{\xi}}{2n^2(m^2\xi-1)}M^2\right)  & \mbox{for } m^2\xi>1. \end{cases}
\end{equation}

Regularizing the above expression via the Riemann's zeta function, $\zeta(s)=\sum_{n=1}^{\infty}n^{-s}$,  we arrive at the Casimir Energy
\begin{equation}\label{vacuumenergy}
E^{(0)}_{reg} \approx \begin{cases} \frac{m^{3}}{8(1-m^2\xi)}M^2   &\mbox {for } m^2\xi<1, \\
\frac{m^{4}\sqrt{\xi}}{8(m^2\xi-1)}M^2 & \mbox{for } m^2\xi>1, \end{cases}
\end{equation}
where we have used the fact that $\zeta(2)=0$ and $\zeta(0)=-1/2$.

On the other side the Casimir tension over the horizon is defined by
\begin{equation}\label{deftension}
\tau_{h}=\frac{\partial E^{(0)}_{reg}}{\partial S_{h}}.
\end{equation}
where $S_h$ is the horizon area. We must be careful since, differently  of the case without RG, now the angular part of the RG metric (\ref{eq:metrica_Kerr-Newman}) also receives a correction and we get
$S_{h}=16\pi M^2g^{-2}(\omega/\omega_P)$. However the remnant Casimir tension is defined just in the limit in which $S_h\sim \mathcal{O}(M^2)$ and we get
\begin{equation}\label{HorizonArea}
S_h \approx \begin{cases} \frac{16 \pi M^2}{(1-m^2\xi)} &\mbox {for } m^2\xi<1, \\
\frac{16n^2\pi(m^2\xi-1)}{m^4\xi}+\frac{16\pi M^2}{m^2\xi-1} & \mbox{for } m^2\xi>1. \end{cases}
\end{equation}
Therefore the horizon area depends on the particle mass, as expected since the metric is coupled to the particle energy.  We should point out that for $m^2\xi<1$ the above expression  reduces to the standard one when we take the limit $\xi\to 0$. For $m^2\xi>1$ this results show us that this model also implies a residual horizon area for the black hole. Now we finally can compute the tension from Eqs. (\ref{deftension}) and (\ref{HorizonArea}). The remnant Casimir tension is defined by $\tau=\lim_{M\to 0}\tau_h$ and we get
\begin{equation}\label{tension}
\tau = \begin{cases} \frac{m^{3}}{128\pi} &\mbox {for } m^2\xi<1, \\
\frac{m^3\sqrt{m^2\xi}}{128\pi} & \mbox{for } m^2\xi>1. \end{cases}
\end{equation}

\begin{figure}[!ht]
 \includegraphics[width=0.8\textwidth]{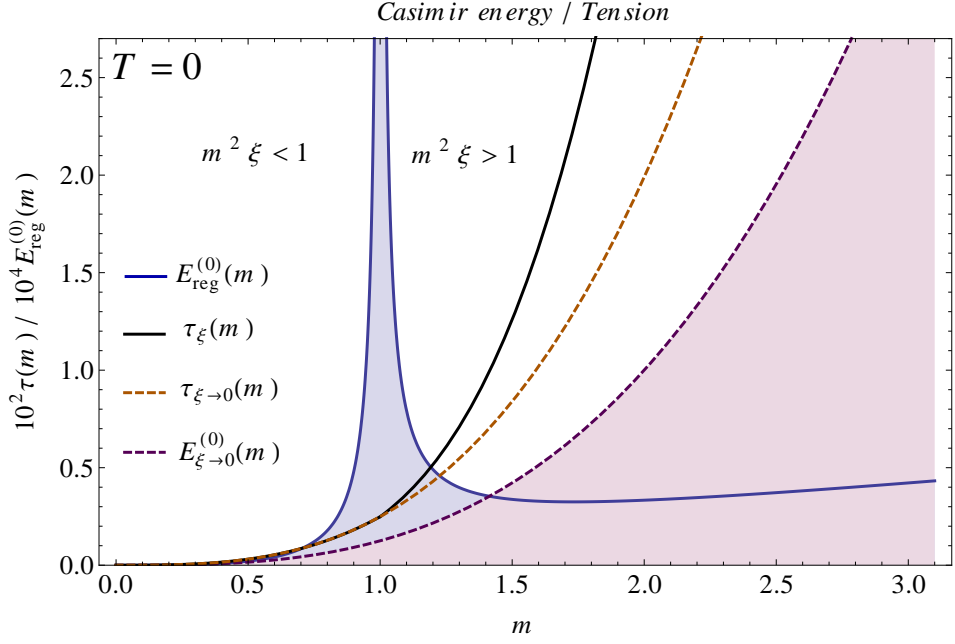}
\caption{Casimir Energy and tension with $M=0.01$ and $\xi=1$.}
  \label{casimir_energy_tension_T0}
\end{figure}

In Fig. (\ref{casimir_energy_tension_T0}) we plot the Casimir energy and the remnant Casimir tension for the cases with and without RG. For the case with RG we can see that despite of the fact that the regularized energy is singular, the tension is well behaved for all values of particle mass. This is due to the fact that the horizon area is also modified and the dependence on $\xi$ is canceled in the computation of the tension. An important point about the tension in the sub-Planckian regime is that it is exactly the value found for the non-deformed Schwarzschild metric analyzed in Ref. \cite{Muniz:2015jba}. This lead us to conclude that the residual tension is an universal property of the black hole (surrounded by a scalar field) evaporation.

\section{Global Casimir Effect in the Rainbow's Gravity at Finite Temperature}
Now we compute the corrections due to the RG to the thermodynamic quantities of the system.  For this we first need of the Helmholtz free energy which is given by \cite{Bordag:2009zzd}
\begin{equation} \label{FreeEnergyHelmholtz}
F^{(0)}=\sum_{n=1}^{\infty}\frac{n^2}{\beta}\log{\left[1-\exp{\left(-\beta  \omega_n\right)}\right]},
\end{equation}
where $\omega_n$ are given by Eq. (\ref{massiveenergies}) and we get
\begin{equation}
F^{(0)}=\sum_{n=1}^{\infty}\frac{n^2}{\beta}\log{\left[1-\exp{\left(-\beta  \left(\frac{m^{2}}{2}+\frac{1}{2\xi}-\frac{\sqrt{m^{4}\xi^{2}-2m^{2}\xi-8m\xi\omega_{n}^{(0)}+1}}{2\xi}\right)^{\frac{1}{2}}\right)}\right]}.
\end{equation}
Since we are interested in the properties of the system when $M\to 0$ we can expand the free energy up to $\mathcal{O}(M^2)$ to get
\begin{equation}
F^{(0)}\approx -\frac{1}{\beta}\sum_{n,k=1}^{\infty}\frac{n^2}{k} e^{-\frac{k\beta}{\sqrt{2\xi}}\sqrt{1+m^2\xi-|1-m^2\xi|}}\left[1+\frac{k\beta m^4\sqrt{\xi} M^2 }{\sqrt{2}n^2|1-m^2\xi|\sqrt{1+m^2\xi-|1-m^2\xi|}} \right].
\end{equation}
Just as the energy in the case at zero temperature, here we get a singular behavior for the free energy at $m\approx \omega_P$. As discussed before we must consider this as a threshold mass pointing to the validity of the system. Anyway we again analyze both regions in order to look for some interesting behavior in the trans-Planckian regime. With this our free energy simplifies to
\begin{equation}
F^{(0)}\approx \begin{cases} -\frac{1}{\beta}\sum_{n,k=1}^{\infty}\frac{n^2}{k} e^{-km\beta}\left[1+\frac{k\beta m^3M^2}{2n^2(1-m^2\xi)} \right] &\mbox {for } m^2\xi<1, \\
-\frac{1}{\beta}\sum_{n,k=1}^{\infty}\frac{n^2}{k} e^{-k\frac{\beta}{\sqrt{\xi}}}\left[1+\frac{k\beta m^4M^2\sqrt{\xi}}{2n^2(m^2\xi-1)} \right] & \mbox{for } m^2\xi>1. \end{cases}
\end{equation}
Now we can use Zeta function regularization of the above expression to get
\begin{equation}\label{FreeEnergyMassive}
F^{(0)}_{reg}\approx \begin{cases} -\sum_{k=1}^{\infty}\frac{\zeta(0) m^3M^2}{2(1-m^2\xi)}e^{-km\beta}
= \frac{1}{(1-m^2\xi)}\frac{ m^3M^2}{4(1-e^{-m\beta})} &\mbox {for } m^2\xi<1, \\
- \sum_{k=1}^{\infty}\frac{\zeta(0) m^4M^2\sqrt{\xi}}{2(m^2\xi-1)}e^{-k\frac{\beta}{\sqrt{\xi}}}
= \frac{ \sqrt{\xi} m^4 M^2 }{4(1-e^{-\frac{\beta}{\sqrt{\xi}}})(m^2\xi-1)} & \mbox{for } m^2\xi>1. \end{cases}
\end{equation}

From the above equation we can get the other thermodynamic quantities. The internal energy is given by
\begin{equation}\label{internalenergy}
U^{(0)}_{reg}(T)=-T^2\frac{\partial{(F^{(0)}_{reg}/T)}}{\partial T}
\end{equation}
and we get
\begin{equation} \label{InternalEnergy}
U^{(0)}_{reg}\approx \begin{cases} \frac{m^3M^2}{1-m^2\xi}\left[\frac{1}{4(1-e^{-\beta m})}-\frac{ \beta m}{16\sinh^2 (\frac{\beta}{2}m)}\right] &\mbox {for } m^2\xi<1, \\
\frac{m^4 M^2}{4(m^2\xi-1)}\left[\frac{ \sqrt{\xi}  }{(1-e^{-\frac{\beta}{\sqrt{\xi}}})}
-\frac{\beta }{4\sinh^2 (\frac{\beta}{2\sqrt{\xi}})}\right] & \mbox{for } m^2\xi>1. \end{cases}
\end{equation}
We should point here that for $m^2\xi<1$ and up to the factor $(1-m^2\xi)^{-1}$, expressions (\ref{FreeEnergyMassive}) and (\ref{InternalEnergy}) are similar to the ones found
in Ref. \cite{Muniz:2015jba} for the case without RG. However their results has some wrong sign which we now correct. As we will comment below, this completely changes the conclusions of Ref. \cite{Muniz:2015jba}. The correct sign can by obtained from our result by taking the limit $\xi \to0$. The same will happens to the remnant Casimir tension.

\begin{figure}[!ht]
     \subfloat[Free Energy with $\beta=1$, $k_B=1$, $M=0.01$, and $\xi=1$.\label{FreeEnergy}]
     {\includegraphics[width=0.48\textwidth]{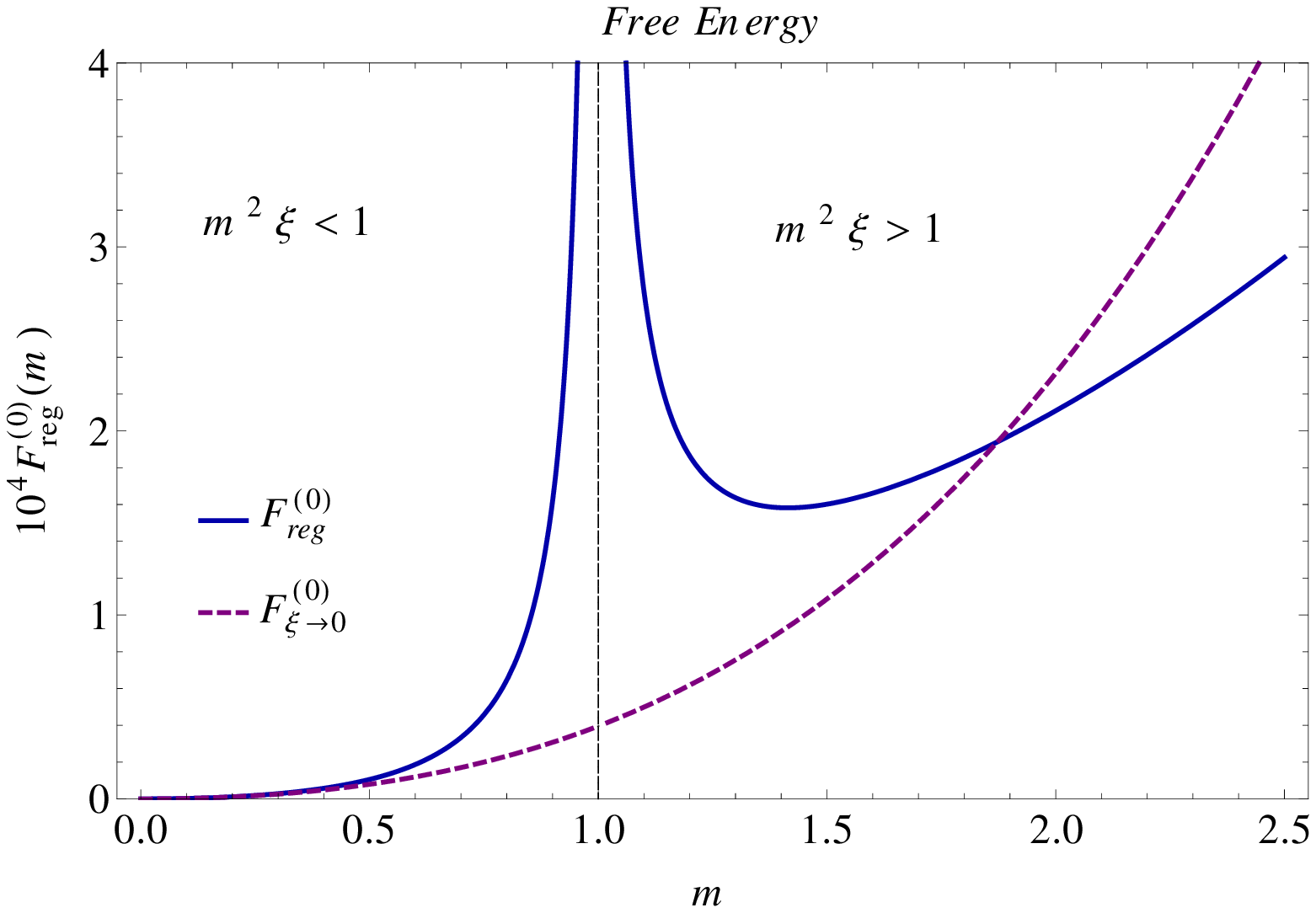}
   }
  \hfill
\subfloat[Heat capacity with $\beta=1$, $M=0.01$, $k_B=1$, and $\xi=1$.\label{HeatCapacity_graf}]
{\includegraphics[width=0.48\textwidth]{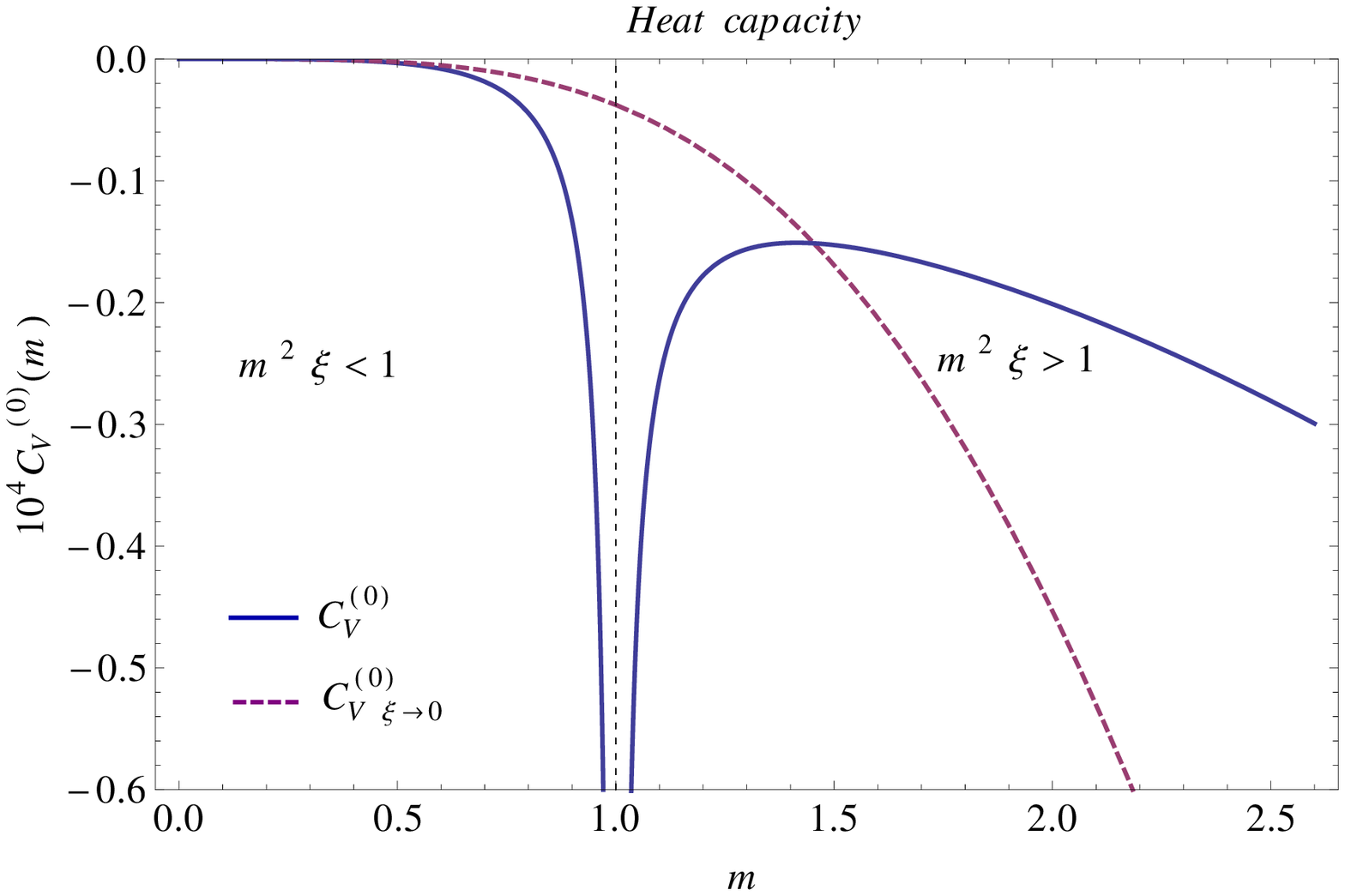}}
\caption{(a) The free energy and (b) heat capacity, for the cases with and without Rainbow's Gravity, depicted in dashed and solid lines respectively.}
\label{FreeEnergyHeatCapacity}
\end{figure}

The entropy and heat capacity are given by
\begin{equation}\label{entropyspecifcheat}
S^{(0)}_{reg}=-\partial F^{(0)}_{reg}/\partial T; C^{(0)}_{_V\,reg}=T\left(\frac{\partial{S^{(0)}_{reg}}}{\partial T}\right)_V .
\end{equation}
The entropy is computed from Eq. (\ref{FreeEnergyMassive})
\begin{equation}\label{Entropy}
S^{(0)}_{reg}\approx \begin{cases} -\frac{ 1}{1-m^2\xi}\frac{k_B\beta^2 m^4 M^2}{16\sinh^2 (\frac{\beta m}{2})} &\mbox {for } m^2\xi<1, \\
-\frac{k_B\beta^2 m^4 M^2}{16(m^2\xi-1)\sinh^2 (\frac{\beta}{2\sqrt{\xi}})} & \mbox{for } m^2\xi>1 \end{cases}
\end{equation}
and from this we get the heat capacity given by
\begin{equation}\label{HeatCapacity_m<1}
C^{(0)}_{_V\,reg}\approx \begin{cases} \frac{k_B m^4 M^2}{8(1-m^2\xi)}\left[ \beta^2\mbox{csch}^2(\frac{\beta m}{2})-\frac{\beta^3 m \mbox{coth}(\frac{\beta m}{2})\mbox{csch}^2(\frac{\beta m}{2})}{2}\right] &\mbox {for } m^2\xi<1, \\
\frac{k_B m^4 M^2}{8(m^2\xi-1)}\left[ \beta^2\mbox{csch}^2(\frac{\beta }{2\sqrt{\xi}})-\frac{\beta^3  \mbox{coth}(\frac{\beta }{2\sqrt{\xi}})\mbox{csch}^2(\frac{\beta }{2\sqrt{\xi}})}{2\sqrt{\xi}}\right] & \mbox{for } m^2\xi>1. \end{cases}
\end{equation}
In Fig. (\ref{FreeEnergyHeatCapacity}) below we plot the free energy and heat capacity of the system. However the free energy is still positive and the heat capacity negative. This means that the quantum vacuum under consideration is thermodynamically unstable for any value of $m$ as in the uncorrected case. Therefore, despite the fact that all thermodynamic quantities are corrected due to RG,  the previous conclusions about stability of the system are basically the same. We stress here that this conclusion is also valid for the trans-Planckian case.

 \begin{figure}[!ht]
     \subfloat[Free Energy with $\beta=1$, $k_B=1$, $M=0.01$, and $m=1$.\label{FreeEnergy_xi}]
     {
      \includegraphics[width=0.49\textwidth]{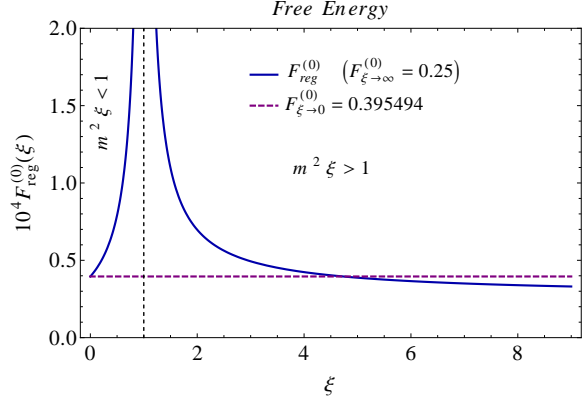}
     }
     \hfill
     \subfloat[Heat capacity with $\beta=1$, $M=0.01$, $k_B=1$, and $ m=1$.\label{free_energy_xi}]
 {
 \includegraphics[width=0.48\textwidth]{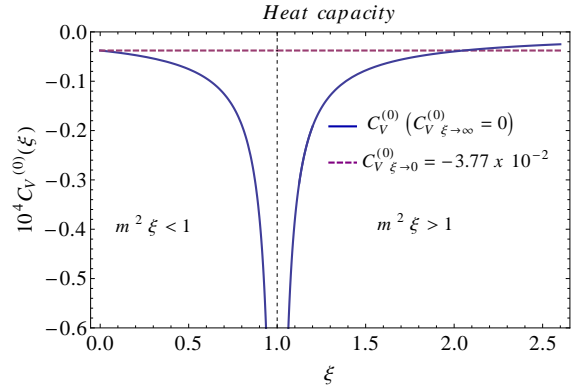}
 }
\caption{(a) The free energy and (b) heat capacity as functions of Rainbow's Gravity parameter $\xi$, compared with the null limit $\xi\rightarrow 0$, depicted in solid and dashed lines respectively. }
   \label{FreeEnergyHeatCapacity_xi}
  \end{figure}

It is also interesting to analyze the high temperature limit of the entropy which is given by

\begin{equation}\label{Entropy High T}
S^{(0)}_{reg}\approx \begin{cases}
-\frac{k_B m^2 M^2}{4(1-m^2\xi)} &\mbox {for } m^2\xi<1, \\
-\frac{k_B\xi m^4 M^2}{4(m^2\xi-1)} & \mbox{for } m^2\xi>1.
\end{cases}
\end{equation}
Therefore, for $m^2\xi<1$ the corrected Casimir entropy is proportional to the horizon area (\ref{HorizonArea}), exactly as in the Hawking entropy. This conclusion was also reached in the case without RG \cite{Muniz:2015jba}.  However, for $m^2\xi>1$, the behavior of the entropy at high temperatures has an anomaly. It is not proportional to the horizon area given in Eq. (\ref{HorizonArea}).

In Fig. (\ref{FreeEnergyHeatCapacity_xi}) we plot the free energy and heat capacity as a function of the Rainbow's Gravity parameter $\xi$ for fixed $m$ and compare with the limit $\xi\rightarrow$ 0. It is worth mentioning that these graphs show,  in the region $m^2\xi>1$, the reduction of the thermodynamic instability of the system for $\xi$ greater than a certain value, in comparison with the $\xi=0$ case. The referred instability  comes from both the positive free energy and negative heat capacity. Such a feature revealed in the graphs of Fig. (\ref{FreeEnergyHeatCapacity_xi}) is entirely compatible with the information contained in the graphs of Fig (\ref{FreeEnergyHeatCapacity}), in which the positive free energy (negative heat capacity) becomes smaller (greater) than the one of the non-deformed black hole from a threshold field mass.

\begin{figure}[!h]
     \includegraphics[width=0.6\textwidth]{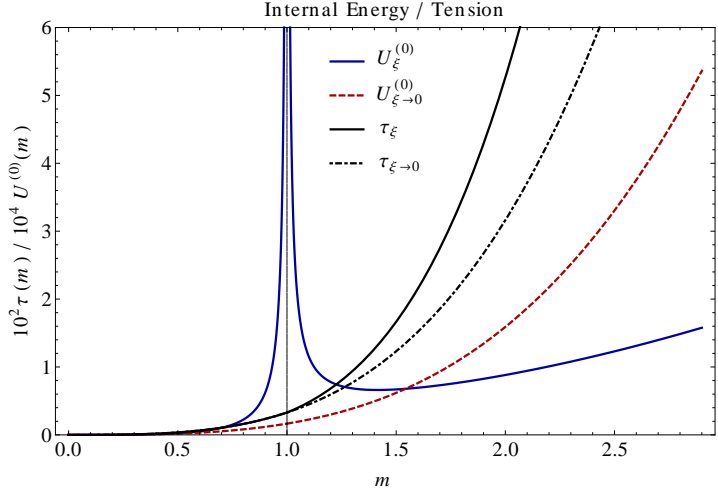}
\caption{Internal Energy with $\beta=1$,$k_B=1$,$M=0.01$ and $\xi=1$.}
 \label{internal_energy}
\end{figure}

Now we turn our attention to the tension over the horizon. This can be computed from Eqs. (\ref{HorizonArea}) and (\ref{InternalEnergy}) to gives us
\begin{equation}\label{tension T}
\tau =\lim _{M\to 0}\frac{\partial U^{(0)}_{reg}}{\partial S_h}= \begin{cases}
\frac{m^3}{64\pi(1-e^{-\beta m})}-\frac{ \beta m^4}{256\pi\sinh^2 (\frac{\beta}{2}m)} &\mbox {for } m^2\xi<1, \\
\frac{  m^3  \sqrt{m^2\xi}}{64\pi(1-e^{-\frac{\beta}{\sqrt{\xi}}})}
-\frac{\beta m^4 }{256\pi\sinh^2 (\frac{\beta}{2\sqrt{\xi}})} & \mbox{for } m^2\xi>1.
\end{cases}
\end{equation}
In Fig. \ref{internal_energy} we plot the internal energy and tension for the finite temperature case. Again we see that despite to the fact the internal energy is singular at $m^2\xi=1$, we have that  the tension is not singular. More important is the fact that for $m^2\xi<1$, as we can from Eq. (\ref{tension T}), the global Casimir effect at finite temperature also do not receive corrections from the RG. We should also analyze the high temperature limit. As said before the internal energy (\ref{InternalEnergy}) do not recover the result of Ref. \cite{Muniz:2015jba} due to some mistakes. The consequence is that the correct remnant tension must also be changed. The correct result is obtained from Eq. \ref{tension T} by taking the limit $\xi \to 0$. This will radically changes the conclusion of that paper for we can see that now the high temperature limit of $\tau$ is given by
\begin{equation}
\lim_{T\to \infty} \tau=m^3/128\pi
\end{equation}

which is exactly the same value of $\tau$ at zero temperature. In Ref. \cite{Muniz:2015jba}  the authors had found a value with opposite sign. By adding this to the tension at $T=0$ the total remnant tension was null. The conclusion was that the remnant tension would be ``erased" at high temperatures. However as exposed above this is not correct and in fact the total tension will be doubled, as compared to the $T=0$ case, when $T\to \infty$. This enforces our conclusion that this is a fundamental and universal property of black hole (surrounded by a scalar field) evaporation.

\section{Closing remarks}

In this manuscript we studied the influence of Rainbow's Gravity (RG) in the global Casimir effect at zero and finite temperature recently investigated by some of the present authors in the general relativity context \cite{Muniz:2015jba}. The actual study consisted of analyzing the stable part of the Casimir energy of a massive scalar field surrounding a Schwarzschild mini-black hole ($M\to 0$) without the presence of material boundaries. The regularization technique was based on the Riemann's zeta function, since the global (or topological) character of the computed quantities dispenses the use of local magnitudes. Despite the fact that the event horizon area and the Casimir quantities vanish when the black hole evaporates completely, it was found that there is a remnant Casimir tension on the horizon, as in the aforementioned work \cite{Muniz:2015jba}.

A natural question that arises is if we can use the Schwarzschild metric to describe the final moments of the black hole evaporation. In order to analyze this, we in fact have considered the effective quantum gravity modification of the Schwarzschild metric described by RG. In this theory the metric depends on the energy of the field in the form given by Eq. (\ref{eq:metrica_Kerr-Newman}), parameterized by a dimensionless quantity $\xi$ and by the Planck's energy $\omega_P$. First we reviewed how to obtain the solutions of the equation of motion for the massive scalar field in this background metric and, in Eq. (\ref{massiveenergies}), found the quantized energies associated with the stationary states around the black hole . This is made in order to investigate the stable part of the regularized quantum vacuum, {\it i.e.}, the one that is neither scattered nor absorbed by the black hole, which provides, therefore, real values for the energies.

The field eigenvalues of energy are given in Eq. (\ref{massiveenergies}) and at this level we already found a singularity at the threshold mass, $m\approx m_P/\sqrt{\xi}$. This suggests that our effective model must be valid only in the sub-Planckian regime $m<m_P/\sqrt{\xi}$. We found that, at zero temperature, both the horizon area and the Casimir energy are modified by a multiplicative factor $(1-m^2\xi/m_P^2)^{-1}$ and therefore are also singular at the threshold mass. In Fig. (\ref{casimir_energy_tension_T0}) we plot the Casimir energy and the remnant Casimir tension. We see that despite the fact that the Casimir energy is singular, the remnant Casimir tension is not. This is due to the fact that the multiplicative factor of the horizon area and energy cancels out. This means in fact that, in the regime $m^2\xi<1$, the tension on the horizon does not receive corrections from RG. This is the first hint that the remnant Casimir tension is an universal property that emerges from complete evaporation of the black hole, which is a consequence of the non-trivial topology of the black hole due to the presence of the singularity at the origin.

Next we considered thermodynamic properties of our system at finite temperature. As in the case at zero temperature, we found in Eqs. (\ref{FreeEnergyMassive}), (\ref{InternalEnergy}), (\ref{Entropy}) and (\ref{HeatCapacity_m<1}) that the free energy, internal energy, entropy and heat capacity are modified by the multiplicative factor $(1-m^2\xi/m_P^2)^{-1}$ in the limit $M\to 0$. Hence we have the same singularity at the field threshold mass. We discover that the result previously found in Ref. \cite{Muniz:2015jba} has some wrong signs and corrected them.  {\color{red} In Fig. (\ref{FreeEnergyHeatCapacity})} we depicted the free energy and heat capacity of the system for $\beta=1$, $M=0.01$, $k_B=1$ and $\xi=1$ in order to analyze thermodynamic stability. We found that for both, the sub-Planckian and the trans-Planckian regimes, the internal energy is positive and the heat capacity is negative. Therefore the system is unstable for any value of the parameter $m$, just as in the case without RG.  We also analyze the high temperature limit of the entropy, Eq. (\ref{Entropy High T}), and found that differently of the case $\xi=0$, an anomaly emerges since the entropy is no more proportional to the horizon area.

Finally we compute the remnant Casimir tension on the horizon at finite temperature in Eq. (\ref{tension T}). In Fig. \ref{internal_energy} we plot the internal energy and tension for $\beta=1$, $M=0.01$, $k_B=1$ and $\xi=1$. We see here that the same behavior as in the $T=0$ case. Namely, despite the fact that the internal energy is singular, the remnant Casimir tension is finite.  Due to the corrections of the signs in the expression for the internal energy, the expression for the remnant tension is also different.  This radically changes the conclusion since now the limit of high temperature of $\tau$ is positive. In fact the value is the same as for the case $T=0$. In Ref. \cite{Muniz:2015jba} it was obtained that the sum of the remnant Casimir tension at $T=0$ and the correction would be zero at high temperatures. However since the sign was wrong in fact the remnant Casimir tension is doubled. We also find that the remnant Casimir tension at finite temperature receives no corrections from RG. Therefore the conclusions are the same as for $\xi=0$ at finite temperature and this enforces our previous suggestion that the remnant Casimir tension can be seen as an universal and fundamental property of the black hole evaporation, and can be related to the spontaneous reduction of the spacetime dimension to $D=2$ which occurs in very high energies according to some effective quantum gravity theories, in the context of black hole physics \cite{Mureika:2012na,Tzikas:2018wzd}, and more recently in the Standard Model of the elementary particles \cite{Addazi:2018xpt}. Thus, the semiclassical approach of our work would point in that direction, since the Casimir tension is fundamentally a linear property, in the sense that it means a force per distance unit. Our results suggest, therefore, that the total evaporation of the black hole would leave behind a kind of one-dimensional scar in the space subject to that tension.

%%%%%%%%%%%%%%%%%%%%%%%%%%%%%%%%%%%%%%%%%%%%%%%%%%%%%%%%%%%%%%%%%%%%%%%%%%%%%%%%%%%%%%%%%%%%%% acknowledgments
%
\section*{Acknowledgements}

The authors would like to thank Alexandra Elbakyan, for removing all barriers in the way of science.
We acknowledge the financial support provided by the Conselho
Nacional de Desenvolvimento Cient\'\i fico e Tecnol\'ogico (CNPq) and Funda\c c\~ao Cearense de
Apoio ao Desenvolvimento Cient\'\i fico e Tecnol\'ogico (FUNCAP) through PRONEM PNE-0112-00085.01.00/16.

%%%%%%%%%%%%%%%%%%%%%%%%%%%%%%%%%%%%%%%%%%%%%%%%%%%%%%%%%%%%%%%%%%%%%%%%%%%%%%%%%%%%%%%%%%%%%%%%%%%%%%%%%%%%%%%%%%%%%%%%%%%%%%%%%%%%%%%%%%%%%%%%%%%%%%%%%%%%%%%%%

\end{document}